\documentclass[journal=jctcce,manuscript=communication]{achemso}
\setkeys{acs}{maxauthors=0,articletitle=true}

\usepackage{achemso}
\usepackage{graphics}
\usepackage{amssymb,amsfonts}
\usepackage{graphicx}
\usepackage[table,dvipsnames]{xcolor}
\usepackage{multirow}
\usepackage{caption}
\usepackage{subcaption}
\usepackage{booktabs}
\usepackage{colortbl}
\usepackage{amsmath}
\usepackage{amsopn}
\usepackage{bm}
\usepackage{siunitx}
\usepackage{bm}
\usepackage{color}
\usepackage{array}
\usepackage{lscape}
\usepackage{mciteplus}
\usepackage[version=3]{mhchem}
\usepackage{ulem}
\usepackage{listings}
\usepackage{enumerate}
\usepackage{lmodern}

\SectionNumbersOn
\DeclareMathOperator{\tr}{Tr}

\author{Janus J. Eriksen}
\email{janus@kemi.dtu.dk}
\affiliation{DTU Chemistry, Technical University of Denmark\\Kemitorvet Bldg. 206, DK--2800 Kgs. Lyngby, Denmark}

\title[TITLE]{Electronic Excitations Through the Prism of Mean-Field Decomposition Techniques}

\begin{document}

%
%
\begin{abstract}

The potential of mean-field decomposition techniques in interpreting electronic transitions in molecules is explored, particularly, the usefulness of these for offering computational signatures of different classes of such excitations. When viewed as a conceptual lens for this purpose, decomposed results are presented for ground- and excited-state energies and dipole moments of selected prototypical organic dyes, and the discrete nature of these properties as well as how they change upon transitioning from one state to another is analyzed without recourse to a discussion based on the involved molecular orbitals. On the basis of results obtained both with and without an account of continuum solvation, our work is further intended to shed new light on practical and pathological differences in between various functional approximations in orbital-optimized Kohn-Sham density functional theory for excited states, equipping practitioners and developers in the field with new probes and possible validation tools.

\end{abstract}

\newpage

%
%

The many successes of the Kohn-Sham formulation of density functional theory (KS-DFT) in computational molecular and materials sciences cannot be disputed nor underestimated by now~\cite{burke_dft_review_jcp_2012,becke_dft_review_jcp_2014,mardirossian_head_gordon_dft_review_mol_phys_2017}, both in its application to ground- and excited-state phenomena in chemistry and physics. For the latter type of problems, the formulation of KS-DFT within linear response theory is commonly referred to as time-dependent DFT (TDDFT)~\cite{casida_tddft_inbook_1995}, although the theory still relies on time-independent exchange-correlation ($xc$) functionals whenever the adiabatic approximation is invoked~\cite{dreuw_head_gordon_tddft_chem_rev_2005}. The popularity of TDDFT notwithstanding, the theory is known to be plagued from a number of severe issues, such as, inaccessibility to states that involve more than a single excitation~\cite{burke_tddft_double_ex_jcp_2004}, core excitations~\cite{besley_asmuruf_tddft_core_ex_pccp_2010}, and charge transfers~\cite{dreuw_head_gordon_long_range_ct_jcp_2003} (CTs). With regards to the difficulties of accounting for the spatial transfer of charge(s) within or between molecules in TDDFT, these are ultimately tied to a wrong account of the electron-transfer self-interaction in the excitation, unlike in time-dependent Hartree-Fock (TDHF) where this contribution is correctly cancelled through the linear response of exact, nonlocal HF exchange~\cite{dreuw_head_gordon_long_range_ct_jacs_2004}. However, KS-DFT may also be directly converged onto arbitrary excited states in a number of alternative ways, even though the overwhelming majority of $xc$ functionals have been designed and parameterized strictly for ground-state purposes. In the arguably most straightforward and intuitive among these, the maximum overlap method (MOM) by Gill and co-workers~\cite{delta_scf_mom_gill_jpca_2008}, the traditional Aufbau principle is discarded in favour of a protocol by which the selection of occupied molecular orbitals (MOs) throughout the self-consistent field (SCF) optimization proceeds via a maximization of the overlap between an updated determinant and that of the previous iteration. As a means to avoid so-called variational collapses back to the lower-lying, stationary Aufbau solution for the ground state, virtual orbitals may be artificially level-shifted to avoid them being filled during the SCF procedure~\cite{saunders_hillier_level_shift_scf_ijqc_1973,carter_fenk_herbert_step_scf_jctc_2020}. For a comprehensive introduction to the revitalized field of orbital-optimized ($\Delta$SCF) approaches to excited-state KS-DFT, the reader is referred to a recent topical perspective by Hait \& Head-Gordon as well as the comprehensive volume of references therein~\cite{hait_head_gordon_oo_dft_jpcl_2021}.\\

The renewed interest in state-specific KS-DFT approaches now begs questions of if, how, and why these methods succeed in capturing the physics of the involved excited states. Particularly, given that the integral Hohenberg-Kohn theorem has no formal counterpart for excited states~\cite{burke_ex_hk_theorem_prl_2004}, we are interested in inspecting and---preferably---quantifying to what extent modern density functional approximations (DFAs) agree on the electronic structure of said states when moving beyond a simple comparison on the basis of what excitation energies these yield and what frontier MOs are involved. In the present study, we will attempt to elucidate any changes to the physics at play in the electronic transition. Particularly, we will seek to measure how perturbations to the electronic structure in an optical transition are manifested through the response of electronic energies and dipole moments to these reorganizations. For this purpose, we will leverage a novel theory for partitioning mean-field KS-DFT first-order properties for a chemical system amongst its constituent atoms, namely, one that decomposes these quantities in an adequate basis of spatially localized MOs~\cite{eriksen_decodense_jcp_2020}. The initial results presented herein collectively show that the use of a localized basis provides a valid new way of analyzing the nature of various types of electronic excitations, without the need to draw inferences from traditional TDDFT-derived metrics based on frontier MOs, e.g., in quantifying and distinguishing between so-called local and CT transitions~\cite{helgaker_tozer_ct_lambda_tddft_jcp_2008,peach_tozer_ct_tddft_jpca_2012,mannucci_adamo_ct_tddft_jctc_2013,dreuw_ct_tddft_jcp_2014,etienne_assfeld_monari_ct_tddft_jctc_2014,kowalski_autschbach_ct_tddft_jctc_2015,krylov_orbital_perspective_jcp_2020}.\\

Recently, our decomposition theory was successfully employed as a means to simulate local properties in condensed phases, with an initial application to liquid, ambient water~\cite{eriksen_local_condensed_phase_jpcl_2021}. In that case, the focused partitioning of solvation energies and dipole moments amongst individual water monomers allowed for the properties to be extracted from truncated water clusters of greatly reduced size. In the present study, by dissecting an electronic-structure simulation on a molecular system and repartitioning the result yielded amongst the constituent atoms, we will instead seek to benefit from the physical soundness of spatially localized MOs to uncover correlations, not in liquids, but in transitions between electronic states, which would otherwise be at risk of staying convoluted in standard mean-field theory.

As outlined in Ref. \citenum{eriksen_decodense_jcp_2020}, a KS-DFT energy may be recast into the following form, partitioned amongst the $\mathcal{M}$ atoms of the system at hand
\begin{align}
E = \sum^{\mathcal{M}}_{K}(E_{\text{elec},K}(\bm{D},\bm{\delta}_K) + E_{xc,K}(\bm{\rho},\bm{\varrho}_K) + E_{\text{nuc},K}) \ , \label{atom_decomp_eq}
\end{align}
in terms of nuclear and electronic contributions defined as
\begin{subequations}
\label{atom_contr_eqs}
\begin{align}
E_{\text{nuc},K} &= Z_K\sum^{\mathcal{M}}_{K>L}\frac{Z_L}{|\bm{r}_K - \bm{r}_L|} \label{nuc_contr_atom_eq} \\
E_{\text{elec},K} &= \tr[\bm{T}_{\text{kin}}\bm{\delta}_K] + \tfrac{1}{2}(\tr[\bm{V}_{K}\bm{D}] + \tr[\bm{V}_{\text{nuc}}\bm{\delta}_{K}]) + \tfrac{1}{2}\sum_{\sigma}\tr[\bm{G}_{\sigma}(\bm{D})\bm{\delta}_{K,\sigma}] \label{elec_contr_atom_eq} \\
E_{xc,K} &= \tr[\epsilon_{xc}(\bm{\rho})\bm{\varrho}_K] \ . \label{xc_contr_atom_eq}
\end{align}
\end{subequations}
In Eq. \ref{nuc_contr_atom_eq}, $Z_K$ and $\bm{r}_K$ denote the nuclear charge and position of atom $K$, while the kinetic energy and nuclear attraction operators in Eq. \ref{elec_contr_atom_eq} are denoted by $\bm{T}_{\text{kin}}$ and $\bm{V}_{\text{nuc}}$, respectively, alongside the attractive potential associated with atom $K$, $\bm{V}_{K}$, and an effective Fock potential, $\bm{G}_{\sigma}$ ($\sigma = \alpha,\beta$ is an electronic spin index). In Eq. \ref{elec_contr_atom_eq}, $\bm{D}$ denotes the full, spin-summed 1-electron reduced density matrix (1-RDM), while the objects that principally define the present decomposition---the atom-specific 1-RDMs, $\{\bm{\delta}\}$---are constructed as follows:
\begin{align}
\bm{\delta}_K &= \sum_{\sigma}\bm{\delta}_{K,\sigma} = \sum_{\sigma}\sum^{\mathcal{N}_{\sigma}}_{i}\bm{d}_{i,\sigma}\bm{p}^{K}_{i,\sigma} \ . \label{atom_rdm1_eq}
\end{align}
In turn, the objects in Eq. \ref{atom_rdm1_eq} are formulated via a set of 1-RDMs, $\bm{d}_{i,\sigma} = \bm{C}_{i,\sigma}\bm{C}^T_{i,\sigma}$, unique to the individual occupied spin-$\sigma$ MOs of the system, $\bm{C}_{i,\sigma}$, and a set of population weights of all $\mathcal{N}_{\sigma}$ MOs of $\alpha$-/$\beta$-spin on a given atom $K$, $\{\bm{p}^{K}\}$. Our earlier investigations in Refs. \citenum{eriksen_decodense_jcp_2020} and \citenum{eriksen_local_condensed_phase_jpcl_2021} have clearly emphasized how the population weights used to assign $\{\bm{d}\}$ should ideally not be drawn from regular Mulliken population analyses~\cite{mulliken_population_jcp_1955}, but rather recast into a basis of reduced dimension, such as, the intrinsic atomic orbitals (IAOs) courtesy of Knizia~\cite{knizia_iao_ibo_jctc_2013,knizia_visscher_iao_jctc_2021}. Finally, the $xc$ energy in Eq. \ref{xc_contr_atom_eq} is expressed in terms of the computed energy density, $\epsilon_{xc}$, as derived from the total electronic density, $\bm{\rho}$, and possibly its derivatives, which are quantities that may be trivially defined in an atom-specific manner, $\{\bm{\varrho}\}$, by proceeding through $\{\bm{\delta}\}$.\\

Eq. \ref{atom_decomp_eq} holds true for any stationary state and we may thus use the theory to decompose transition energies, ${\Delta}E$, whenever the excited state in question has been obtained in an orbital-optimized manner. For vertical Frank-Condon transitions, ${\Delta}E_{\text{nuc},K} = 0 \ \forall \ K$, and we are left with sets of atom-specific, electronic contributions associated with each of the operators in Eqs. \ref{elec_contr_atom_eq} and \ref{xc_contr_atom_eq}. Specifically, we will distinguish between these by grouping them as follows: ($\bm{i}$) kinetics, ($\bm{ii}$) nuclear attraction, ($\bm{iii}$) Coulombic repulsion, ($\bm{iv}$) exact exchange, and ($\bm{v}$) KS exchange-correlation (cf. the individual operators in Eqs. \ref{elec_contr_atom_eq} and \ref{xc_contr_atom_eq}). In addition, we have here further extended our theory to allow for mean-field decompositions in the presence of domain-decomposed continuum solvation models~\cite{cances_maday_stamm_dd_solvation_jcp_2013}, which may generally be formulated as a grid-based summation over the atomic electrostatic solvation energies, each of which is a product of the total 1-RDM and its potential. We will herein present results obtained using the ddCOSMO model~\cite{lipparini_mennucci_dd_cosmo_jctc_2014,klamt_cosmo_review_wires_2017}; besides adding specific electrostatic energy contributions, the solvation model further implicitly alters all other contributions as well.\\

The relevant contributions in Eq. \ref{elec_contr_atom_eq} have been formulated in terms of standard Coulomb, $\bm{J}$, and exchange, $\bm{K}$, integrals, where the exchange ratio is $0 < \alpha$ only for $xc$ functionals at the hybrid level or higher~\cite{perdew_jacobs_ladder_aip_conf_proc_2001}. Barring the solvation energy and the second term in Eq. \ref{elec_contr_atom_eq}, all other individual contributions depend directly on the atom-specific 1-RDMs in Eq. \ref{atom_rdm1_eq}. In many other alternative decomposition schemes, particularly those that are similarly exact (i.e., lossless), properties are instead partitioned amongst the constituent atoms on the basis of the full 1-RDM. In the arguably most intuitive example of such a decomposition, the energy density analysis (EDA) by Nakai~\cite{nakai_eda_partitioning_cpl_2002,nakai_eda_partitioning_ijqc_2009}, $\bm{D}$ is partitioned on account of which atoms the individual AOs are localized on (that is, irrespective of any further population measures) by limiting the trace operations for most of the products with $\bm{D}$ in Eqs. \ref{atom_contr_eqs} to only those AOs that are assigned to atom $K$. However, one term, namely, the first contribution to the electron-nuclear attraction in Eq. \ref{elec_contr_atom_eq}, remains shared between these two decomposition schemes. Denoting this term as a {\textit{global}} nuclear attraction (since it depends on the full, orbital-invariant 1-RDM, on par with the solvation energy), all other electronic contributions will generally differ in between the present and the EDA partitioning, including the latter of the two nuclear-attraction terms, which we will henceforth denote as {\textit{local}}, given that the involved 1-RDM object is atom-centric rather than defined for the full molecule~\bibnote{Please note that the sum of the local and global electron-nuclear attractions will be identical, whereas this will in general not hold true in the case of the individual atomic contributions.}.\\

\begin{figure}[ht!]
\vspace{-.6cm}
\begin{center}
\includegraphics[width=.85\textwidth]{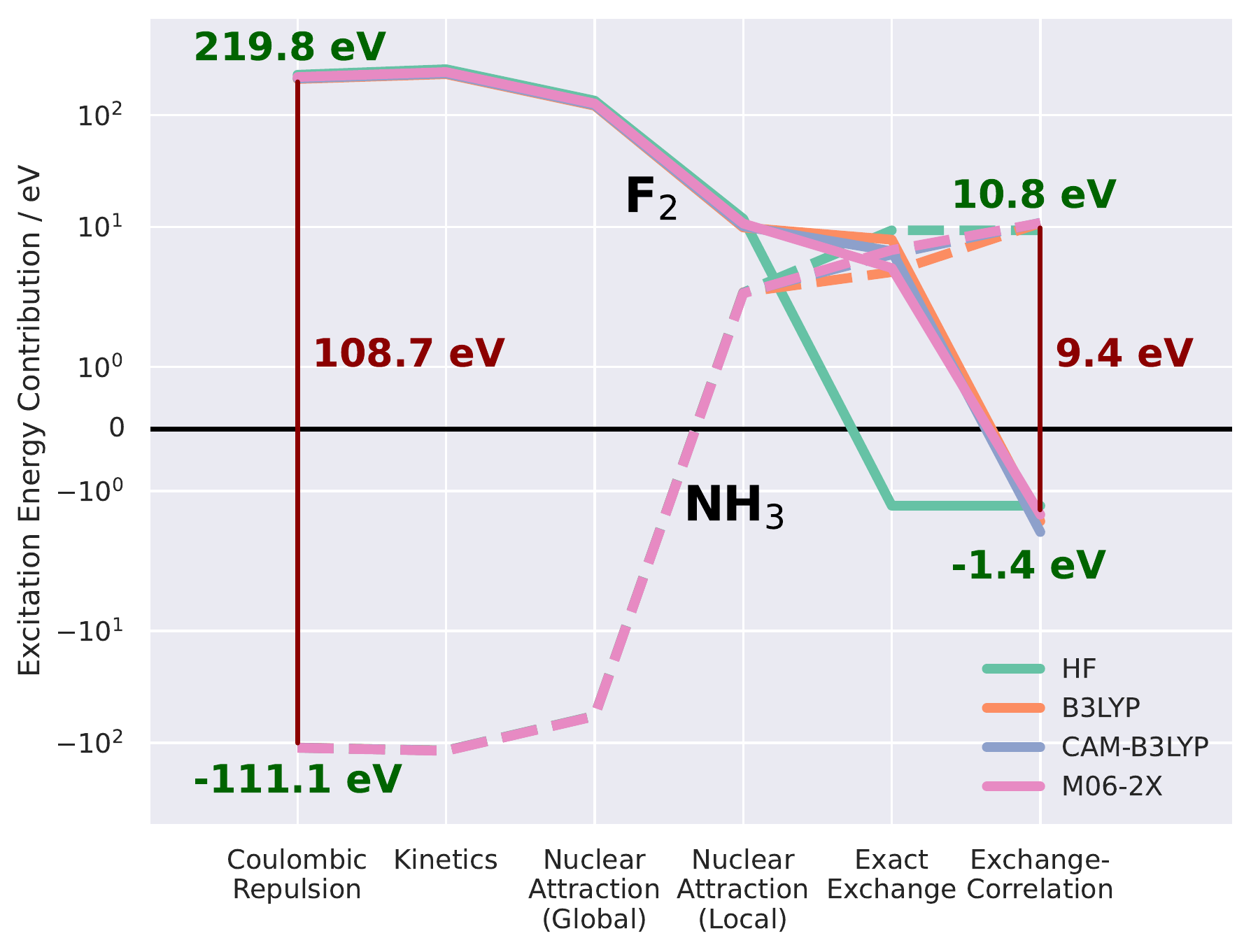}
\caption{Accumulated contributions to the total excitation energy of the CT transition in the separated F$_2$ $\cdots$ NH$_3$ system, as calculated using the present decomposition, HF alongside a selection of DFAs, the aug-pc-1 basis set, and a combination of IBOs/IAOs. Selected individual contributions for F$_2$ (solid lines) and NH$_3$ (dashed lines) using the M06-2X functional are indicated in green, while the sum of these are annotated in dark red. Please see Table S1 of the Supporting Information for all individual contributions.}
\label{fig_1}
\end{center}
\vspace{-.8cm}
\end{figure}
To being with, we study the pathological CT transition recently featured in Ref. \citenum{hait_head_gordon_oo_dft_jpcl_2021} between a pair of isolated F$_2$ and NH$_3$ molecules, separated at a distance of 1000 \AA, leading to a supersystem consisting of charged $\text{F}^{-}_2$ and $\text{NH}^{+}_3$ fragments. Throughout our study, the DFAs employed are B3LYP~\cite{becke_b3lyp_functional_jcp_1993,frisch_b3lyp_functional_jpc_1994}, CAM-B3LYP~\cite{yanai_tew_handy_camb3lyp_functional_cpl_2004}, and M06-2X~\cite{zhao_truhlar_m06_functional_tca_2008} (in addition to standard HF) in the aug-pc-$n$ family of (augmented) double-, triple-, and quadruple-$\zeta$ quality basis sets ($n =$ 1, 2,  3)~\cite{jensen_pc_basis_sets_jcp_2001}. The unrestricted $\Delta$SCF results for the lowest-lying CT triplet state in the separated F$_2$ $\cdots$ NH$_3$ system are presented in Fig. \ref{fig_1}, partitioned into accumulated contributions after the inclusion of each of the non-vanishing operator terms in Eqs. \ref{atom_contr_eqs}. As is clear by traversing through the individual contributions in Fig. \ref{fig_1}, the accumulated contributions from, e.g., exact exchange and exchange-correlation effects may appear minor on the whole (monitoring only total changes to the excitation energy, the sum of the F$_2$ and NH$_3$ contributions changes by a mere $-2.8$ eV and an additional $-1.0$ eV, respectively, in the case of M06-2X), but this need not necessarily be true in the case of what alterations to the underlying atomic contributions these produce. Most pronouncedly, while the inclusion of $xc$ contributions from Eq. \ref{xc_contr_atom_eq} is observed to only lower excitation energies by $1-2.5$ eV depending on the mean-field treatment of choice, it radically shifts the balance between positive and negative atomic contributions to these. In uncorrelated HF theory, this shift is assigned in full to the inclusion of exact exchange on top of all other contributions.\\

Looking at the total contributions to the excitation energy associated with each of the two molecules, the electronic structure of F$_2$ is observed to be only marginally perturbed ($E_{\text{F}_2} = -1.382$ eV), whereas the opposite is true for the NH$_3$ molecule, to which the partitioned excitation energy is almost exclusively assigned ($E_{\text{NH}_3} = 10.818$ eV). This observation aligns well with chemical intuition as the true excitation energy should be equal to the sum of the signed vertical ionization potential of NH$_3$ and electron affinity of F$_2$, which are calculated to be $E_{\text{IP}(\text{NH}_3)} = 10.826$ eV and $E_{\text{EA}(\text{F}_2)} = -1.375$ eV at the M06-2X/aug-pc-1 level of theory. The consistency by which atomic contributions derived from the present decompositions align with thermochemical data is thus assuring, in the same way as was true for the agreement between the simulated solvation energy of a single water monomer in the bulk phase in Ref. \citenum{eriksen_local_condensed_phase_jpcl_2021} and the tabulated enthalpy of vaporization of liquid water at \SI{25}{\celsius}. Due to point-group symmetry, the EDA partitioning yields identical results for F$_2$ in this case, but not NH$_3$, for which the individual contributions to the excitation energy show the effect of the positive partial charge to be almost evenly distributed among the N and H atoms, whereas the present decomposition localizes this on the nitrogen to a greater extent.\\

\begin{figure}[ht!]
\vspace{-.6cm}
\begin{center}
\includegraphics[width=.83\textwidth]{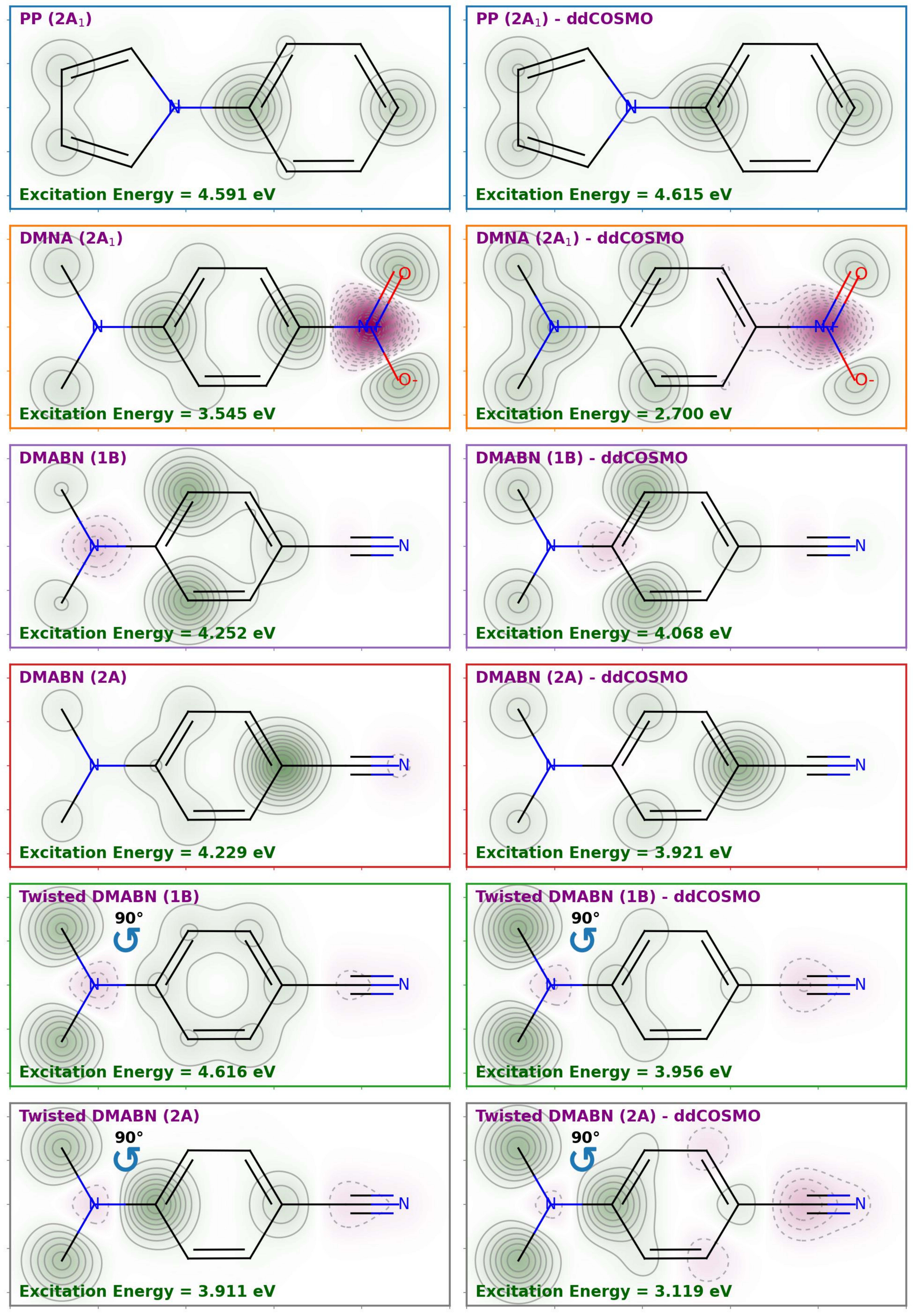}
\caption{Total changes in individual atomic electronic energy contributions along the studied transitions in PP, DMNA, and (twisted) DMABN (in vacuo or ddCOSMO solution), as calculated using Eqs. \ref{atom_contr_eqs} at the CAM-B3LYP/aug-pc-1 level of theory and a combination of IBOs/IAOs. In all plots, pink and green colors indicate negative and positive shifts, respectively, and the plots have been normalized across all individual transitions.}
\label{fig_2}
\end{center}
\vspace{-.8cm}
\end{figure}
Results for excitation energies across a selection of prototypical aromatic donor-acceptor systems are next presented in Fig. \ref{fig_2}, with the magnitude and sign of the atomic decompositions of the excitation energies depicted by superimposing Gaussian distributions of these onto molecular 2D structures generated using {\texttt{RDKit}~\cite{rdkit_prog}. Please note that, for the sake of brevity, all contributions associated with hydrogen atoms have been carefully folded onto their neighbouring heavy atoms. The transitions in Fig. \ref{fig_2} have all previously been established as either local or CT in the literature~\cite{luethi_donor_accept_tddft_jcp_2002,furche_dmabn_tddft_jacs_2004}; however, these classifications have all relied on various diagnostics, most of which are based on traditional MO-based (TDDFT-derived) inference maps. The systems in Fig. \ref{fig_2} are abbreviated as follows: {\textit{N}}-phenylpyrrole (PP), {\textit{N,N}}-dimethyl-4-nitroaniline (DMNA), and 4-(dimethylamino)benzonitrile (DMABN). In the latter case, results have been computed for both a planar and a twisted (by $90^{\circ}$) configuration to mimic the two limits in its twisted intramolecular CT process. In all cases, optimized gas-phase geometries have been borrowed from the recent benchmark study in Ref. \citenum{loos_jacquemin_cipsi_ct_state_jctc_2021}.\\

Fig. \ref{fig_2} shows results for unrestricted $\Delta$SCF simulations of the singlet states in question, while Fig. S1 of the Supporting Information (SI) presents corresponding results subject to approximate spin-projection (i.e., implicitly relying on simulations of the corresponding triplet states as well) in an attempt to remove any possible spin contamination~\cite{houk_approx_spin_proj_cpl_1988}. As was the case in Fig. \ref{fig_1}, no recourse to the concept of frontier orbital pairs has been made, except for fixing the occupation of the excited state. Besides the use of IAOs for estimating the charge populations to be used in Eq. \ref{atom_rdm1_eq}, these results (along with those in Fig. \ref{fig_1}) have been obtained in a basis of standard intrinsic bond orbitals~\cite{knizia_iao_ibo_jctc_2013} (IBOs) with a separate localization of core and valence MOs~\bibnote{As in our previous studies (Refs. \citenum{eriksen_decodense_jcp_2020} and \citenum{eriksen_local_condensed_phase_jpcl_2021}), a PM localization power ($p=2$) has been used to generate the IBOs~\cite{lehtola_jonsson_pm_jctc_2014}.}. Results obtained using other types of spatially localized orbitals, i.e., Foster-Boys~\cite{foster_boys_rev_mod_phys_1960} (FB) and Pipek-Mezey~\cite{pipek_mezey_jcp_1989} (PM), are collected in Fig. S2 of the SI. In summary, both these types of MOs are found to be clearly inferior to IBOs for the present purpose, e.g., by yielding results that fail to reflect the underlying point-group symmetry used, on par with what we have previously observed in both of our earlier studies~\cite{eriksen_decodense_jcp_2020,eriksen_local_condensed_phase_jpcl_2021}.\\

\begin{figure}[ht!]
\vspace{-.6cm}
\begin{center}
\includegraphics[width=.96\textwidth]{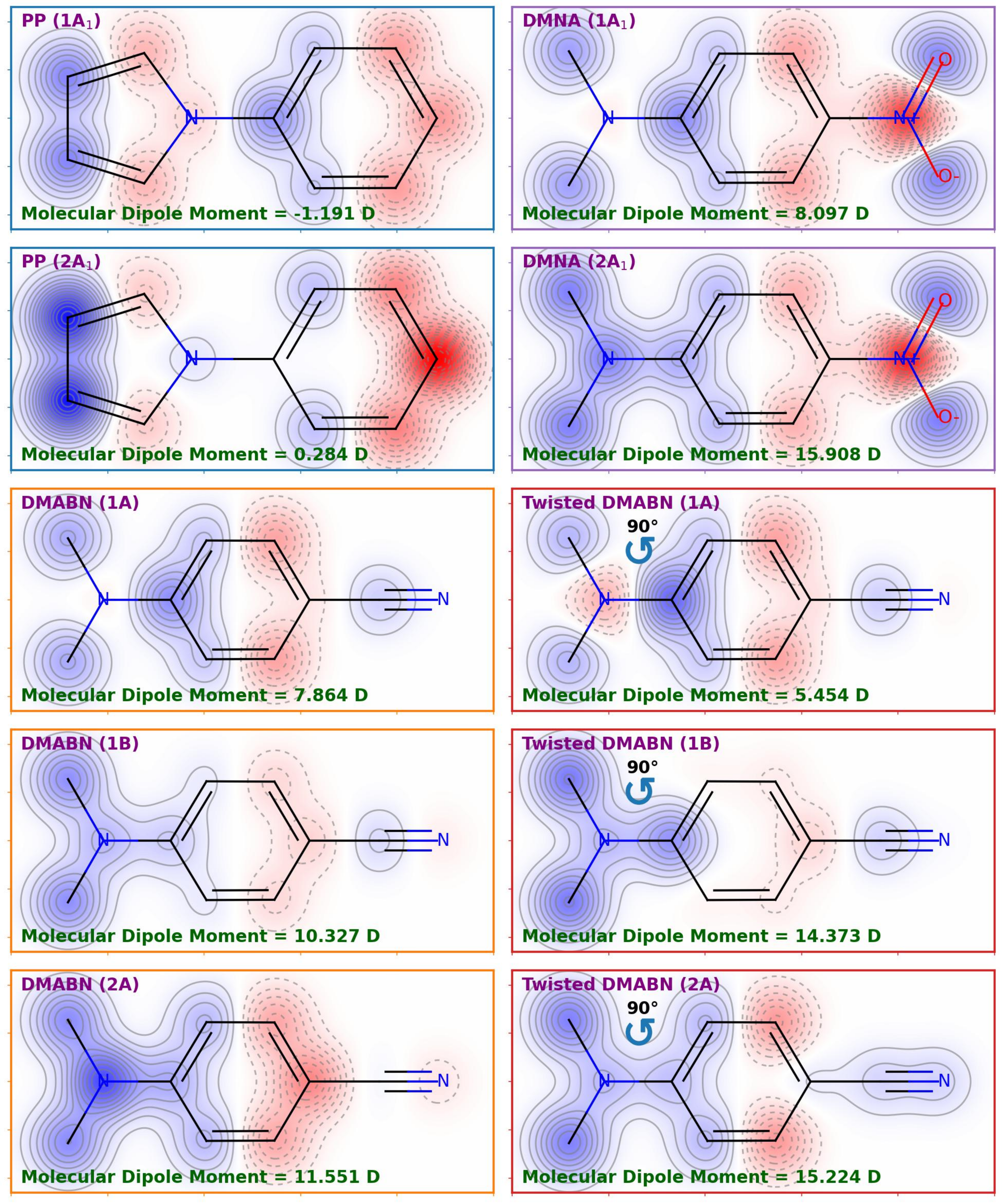}
\caption{Individual atomic contributions to the molecular dipole moment (in vacuo) of each of the studied singlet states in PP, DMNA, and (twisted) DMABN, as calculated using Eqs. \ref{atom_contr_dipmom_eqs} at the CAM-B3LYP/aug-pc-1 level of theory and a combination of IBOs/IAOs. In all plots, red and blue colors indicate negative and positive contributions along the longitudinal $z$-axis (chemistry convention, with the positive direction pointing from donor to acceptor), respectively, and the plots have once again been normalized across all states.}
\label{fig_3}
\end{center}
\vspace{-.8cm}
\end{figure}
In Fig. \ref{fig_3}, we further complement the results in Fig. \ref{fig_2} by providing an alternative visualization of possible charge movements by means of corresponding decompositions of the molecular dipole moments of the states in question. For that purpose, we note that dipole moments---irrespective of the level of mean-field theory used---may be expressed as
\begin{align}
\bm{\mu}_{\text{MF}} = \sum^{\mathcal{M}}_{K}(\bm{\mu}_{\text{elec},K}(\bm{\delta}_K) + \bm{\mu}_{\text{nuc},K}) \ , \label{atom_decomp_dipmom_eq}
\end{align}
with nuclear and electronic contributions defined as
\begin{subequations}
\label{atom_contr_dipmom_eqs}
\begin{align}
\bm{\mu}^{\alpha}_{\text{nuc},K} &= Z_{K}\bm{r}^{\alpha}_{K} \label{nuc_contr_atom_dipmom_eq} \\
\bm{\mu}^{\alpha}_{\text{elec},K} &= -\tr[\bar{\bm{\mu}}^{\alpha}\bm{\delta}_{K}] \ , \label{elec_contr_atom_dipmom_eq}
\end{align}
\end{subequations}
in terms of AO dipole integrals, $\bar{\bm{\mu}}^{\alpha}$, for each of the Cartesian components, $\alpha=x,y,z$. As such, Fig. \ref{fig_3} seeks to measure the depletion (augmentation) of electronic density along the transitions in Fig. \ref{fig_2} in a more rigorous manner than if we had opted to monitor changes to partial atomic charges, as such results are orbital-invariant (relying on the total 1-RDM).\\

We begin by looking at the results for the PP(2A$_1$) transition in Figs. \ref{fig_2} and \ref{fig_3}. Operating under the premise that this excitation is indeed of entirely local $\pi$-$\pi^{\ast}$ character (whatever the interpretation of this entails when we move beyond an MO picture), we observe how all atomic energy contributions in Fig. \ref{fig_2} are uniformly positive in this case, in vacuo as well as solution. However, the contributions to the total excitation energy are hardly localized on just one or a pair of centers, making any parallel to be drawn to the well-established interpretation from the classical orbital picture, e.g., as a $\pi$-$\pi^{\ast}$ excitation from the nitrogen lone pair to the phenyl moiety, somewhat difficult as the overall distributions in general appear more composite altogether. From the decomposed dipole moments in Fig. \ref{fig_3}, the change in direction of the molecular dipole moment is seen to result from an increase in polarity, although the charge distributions appear very similar in the ground and excited states.\\

In Fig. S7 of the SI, the PP results of Fig. \ref{fig_2} are partitioned into individual contributions from each of the operators in the Hamiltonian, akin to Fig. \ref{fig_1}. As discussed earlier, the total, accumulated results in Fig. \ref{fig_2} come about in complex manners, with each of the operator terms in Eq. \ref{atom_contr_eqs} contributing in various, system-dependent manners, but we satisfactorily note that all three of the tested DFAs are observed to agree on these distributions (cf. Fig. S3). The same holds true with respect to the employed one-electron basis set; Fig. S4 of the SI convincingly demonstrates how not only the total excitation energies hardly vary upon extending the aug-pc-$n$ basis sets, for each and every of the studied transitions, but likewise for the individual atomic contributions. In general, all tested DFAs agree on the impact of the individual, decomposed components to the total excitation energies, and differences in between the $xc$ functionals thus result from varying absolute magnitudes of each of these. The reason behind this similarity is the resemblance of the spatially localized MOs yielded by the different DFAs, as also confirmed by comparing total excitation energies in Fig. S5 of the SI for the six states in question obtained with the B3LYP and CAM-B3LYP $xc$ functionals, both excluding and including ddCOSMO solvent effects. In Fig. S5, these results are further compared to corresponding TDDFT results, which show much larger variances, sensitive to both the DFA used and in the magnitude of the solvatochromic shift~\cite{hait_head_gordon_oo_dft_jpcl_2021}.\\

In the following panel of Fig. \ref{fig_2} (as well as Fig. S8 of the SI), results are reported for the transition to the 2A$_1$ state in DMNA, which by all possible metrics is understood to involve charge transferred from the (donor) amino group to the (acceptor) nitro group across the aromatic linker. In addition, Fig. S9 of the SI presents corresponding results for the closely related 4-nitroaniline (NA) system, for which it is fair to expect decomposed results to resemble those for the DMNA(2A$_1$) transition, while at the same time differing from those for the above PP(2A$_1$) transition. Traditionally, both transitions to the 2A$_1$ states have been interpreted as being accompanied by a predominant shift from the canonical Lewis structure depicted in Fig. \ref{fig_2} to the corresponding zwitterionic resonance structure, particularly in a polar solvent, as discussed at length in, e.g., Ref. \citenum{eriksen_pna_tddft_mol_phys_2013}. Comparing the results for the DMNA(2A$_1$) and PNA(2A$_1$) transitions in Figs. S8 and S9, respectively, these do indeed bear a strong and convincing resemblance. This is, however, not even remotely true for the alternative EDA partitioning, cf. Fig. S6 of the SI, which remains predetermined by (while scaling with) the composition and extent of the underlying basis set due to the dependence on $\bm{D}$. In Fig. \ref{fig_2}, the DMNA(2A$_1$) results show atoms, for which the associated energy contributions are increased in the transition to the excited state (on par with the PP(2A$_1$) results), but also atoms, which energetically benefit from this photoinduced process. Given what is known about this transition above all certainty, we note that our decomposition---in contrast to the transition in PP---yields a collection of contributions of opposite signs, which we will henceforth proceed to interpret as a possible signature of CT, at least for the present type of separated donor-acceptor systems. This hypothesis is further supported by the fact that more atoms get assigned a negative contribution to the excitation energy in a polar environment than in vacuo. Lastly, in terms of the decomposed dipole moment in Fig. \ref{fig_3}, the difference in the strength of the individual contributions as well as the overall polarity around the amino group in the ground and excited states are significantly more pronounced than what was observed in the case of PP, in agreement with the expected shift to a distinctly more charge-separated electronic structure in the 2A$_1$ state.\\

In the final four panels of Fig. \ref{fig_2}, results are presented for transitions to the two lowest-lying excited states in planar and twisted DMABN (cf. also individual results in Figs. S10--S13 of the SI). The two excited states in question are denoted by A and B term symbols herein, since they transform according to these irreducible representations of the $C_2$ point group and this then allows for them to be identified across all values of the twist angle leading from the planar to the twisted geometry. As detailed in Ref. \citenum{grabowski_rettig_dmabn_chem_rev_2003}, the donor and acceptor units effectively become electronically uncoupled for this system upon enforcing orthogonality between the amino group and the phenyl moiety in the twisted configuration. In particular, the 2A excited state, which has commonly been referred to as the CT state among the two, is believed to relax into a highly dipolar quinoidal electronic structure, resulting in a characteristically enlarged dipole moment, cf. the results in Fig. \ref{fig_3}.\\

However, assessing only the length of the total molecular dipole moment as an indicator of a charge-separated excited state is obviously too simplified a distinction, given that the 1B and 2A states fail to differ much from one another based on this metric alone. Comparing instead the atomic partitioning of these dipole moments is observed from Fig. \ref{fig_3} to reveal a more detailed picture. Whereas both the distributions of atomic contributions to the ground (1A) and excited (1B) state dipole moments are enhanced upon twisting the amino group out of the plane, the distribution of the dipole moment for the 2A state is moreover seen to change sign in the contributions assigned to the terminal nitrile group. However, both excited states are obviously polarized in their electronic structures and considerably so in the twisted configuration. This is further supported by inspecting the decomposed excitation energies in Fig. \ref{fig_2}. As noted earlier, results that span a collection of atomic contributions of opposite, rather than equal signs may be indicative of a proper CT process in between a pair of units that donate and accept electrons, respectively, but this remains a measure which need obviously be evaluated in combination with the length scale at which these separations appear. As such, neither of the two excited states in the planar configuration appears particularly charge-separated, which is further reflected by the weak solvatochromic shift of both in moving from in vacuo to solution. In the twisted configuration, however, the distributions of atomic contributions appear more separated and the solvent responses are likewise significantly stronger. In future studies, it might prove instructive to monitor the changes to both sets of decomposed properties for these two states in a finer sampling along the twist angle, but this unfortunately lies outside the scope of the present work.\\

The present study has proposed and numerically demonstrated a new conceptual lens through which to analyze, quantitatively assess, and differentiate various kinds of electronic transitions, without recourse to traditional MO-based inference maps in the form of population analyses, natural transition orbitals, and attachment/detachment densities. In summary, our results appear to show that a collection of atomic energy contributions of opposite signs, rather than equal, may in itself serve as an indication of a CT excitation in standard donor-acceptor systems, although the spatial separation of these units necessarily needs to be taken into account as well. Likewise, decomposed molecular dipole moments, and how these change in transitioning from one state to another, may aid in elaborating and further refining upon current standardized definitions of what local and CT excitations amount to by viewing these within an entirely different theoretical frame. It is our hope that these signatures may aid in paving the way towards a more sophisticated understanding of the physics at play in a range of complex optoelectronic processes taking place across extended $\pi$-conjugated networks, with a significant relevance in contemporary materials sciences and beyond. Among possible future applications, the present decompositions may shed new light on various electroluminescent processes, e.g., thermally activated delayed fluorescence for indirectly harvesting triplet excitons in OLEDs~\cite{adachi_tadf_nature_2012,hait_voorhis_tadf_jctc_2016,de_silva_tadf_jpcl_2019}, or the intramolecular singlet fission process at the heart of new classes of emerging organic solar cells~\cite{casanova_sf_chem_rev_2018,corminboeuf_sf_jpcl_2020}. Besides MOM-based excited-state simulations, constrained KS-DFT techniques offer yet another means by which we may probe and decompose these phenomena~\cite{van_voorhis_const_dft_chem_rev_2012}. Finally, the fact that fundamentally unrelated DFAs have here been found to yield decomposed results in qualitative agreement with one another, but with subtle differences in the magnitudes of each of the studied components, prompts the intriguing question as to whether these responses might serve a possible practical purpose in the inverse engineering of future $xc$ functionals.

%
%
\section*{Acknowledgments}

This work was supported by a research grant (no. 37411) from VILLUM FONDEN (a part of THE VELUX FOUNDATIONS). The author thanks Filippo Lipparini (Universit{\`a} di Pisa) and J{\"u}rgen Gauss (Johannes Gutenberg-Universit{\"a}t Mainz) for providing useful comments.

%
%
\section*{Supporting Information}

The Supporting Information collects a number of additional results. Table S1 lists the individual contributions behind Fig. 1. Fig. S1 shows the effect of approximate spin-projection on decomposed excitation energies, and Figs. S2--S4 compare different orbital localization schemes, DFAs, and basis sets of increasing size for the same set of systems, respectively. Fig. S5 compares total TDDFT and $\Delta$SCF excitation energies, while Fig. S6 compares the present set of results to corresponding decompositions calculated using Nakai's EDA technique. Finally, Figs. S7 through S13 compare the individual contributions to the total transition energies for each of the studied systems in Fig. \ref{fig_2}.\\

\noindent The \texttt{decodense} code was used to perform all KS-DFT decompositions:\\
{\url{https://github.com/januseriksen/decodense}}

\providecommand{\latin}[1]{#1}
\makeatletter
\providecommand{\doi}
  {\begingroup\let\do\@makeother\dospecials
  \catcode`\{=1 \catcode`\}=2 \doi@aux}
\providecommand{\doi@aux}[1]{\endgroup\texttt{#1}}
\makeatother
\providecommand*\mcitethebibliography{\thebibliography}
\csname @ifundefined\endcsname{endmcitethebibliography}
  {\let\endmcitethebibliography\endthebibliography}{}


\begin{mcitethebibliography}{54}
\providecommand*\natexlab[1]{#1}
\providecommand*\mciteSetBstSublistMode[1]{}
\providecommand*\mciteSetBstMaxWidthForm[2]{}
\providecommand*\mciteBstWouldAddEndPuncttrue
  {\def\EndOfBibitem{\unskip.}}
\providecommand*\mciteBstWouldAddEndPunctfalse
  {\let\EndOfBibitem\relax}
\providecommand*\mciteSetBstMidEndSepPunct[3]{}
\providecommand*\mciteSetBstSublistLabelBeginEnd[3]{}
\providecommand*\EndOfBibitem{}
\mciteSetBstSublistMode{f}
\mciteSetBstMaxWidthForm{subitem}{(\alph{mcitesubitemcount})}
\mciteSetBstSublistLabelBeginEnd
  {\mcitemaxwidthsubitemform\space}
  {\relax}
  {\relax}

\bibitem[Burke(2012)]{burke_dft_review_jcp_2012}
Burke,~K. {Perspective on Density Functional Theory}. \emph{J. Chem. Phys.}
  \textbf{2012}, \emph{136}, 150901\relax
\mciteBstWouldAddEndPuncttrue
\mciteSetBstMidEndSepPunct{\mcitedefaultmidpunct}
{\mcitedefaultendpunct}{\mcitedefaultseppunct}\relax
\EndOfBibitem
\bibitem[Becke(2014)]{becke_dft_review_jcp_2014}
Becke,~A.~D. {Perspective: Fifty Years of Density-Functional Theory in Chemical
  Physics}. \emph{J. Chem. Phys.} \textbf{2014}, \emph{140}, 18A301\relax
\mciteBstWouldAddEndPuncttrue
\mciteSetBstMidEndSepPunct{\mcitedefaultmidpunct}
{\mcitedefaultendpunct}{\mcitedefaultseppunct}\relax
\EndOfBibitem
\bibitem[Mardirossian and Head-Gordon(2017)Mardirossian, and
  Head-Gordon]{mardirossian_head_gordon_dft_review_mol_phys_2017}
Mardirossian,~N.; Head-Gordon,~M. {Thirty Years of Density Functional Theory in
  Computational Chemistry: An Overview and Extensive Assessment of 200 Density
  Functionals}. \emph{Mol. Phys.} \textbf{2017}, \emph{115}, 2315\relax
\mciteBstWouldAddEndPuncttrue
\mciteSetBstMidEndSepPunct{\mcitedefaultmidpunct}
{\mcitedefaultendpunct}{\mcitedefaultseppunct}\relax
\EndOfBibitem
\bibitem[Casida(1995)]{casida_tddft_inbook_1995}
Casida,~M.~E. \emph{{Recent Advances in Density Functional Methods}}; World
  Scientific, 1995\relax
\mciteBstWouldAddEndPuncttrue
\mciteSetBstMidEndSepPunct{\mcitedefaultmidpunct}
{\mcitedefaultendpunct}{\mcitedefaultseppunct}\relax
\EndOfBibitem
\bibitem[Dreuw and Head-Gordon(2005)Dreuw, and
  Head-Gordon]{dreuw_head_gordon_tddft_chem_rev_2005}
Dreuw,~A.; Head-Gordon,~M. {Single-Reference {\textit{Ab Initio}} Methods for
  the Calculation of Excited States of Large Molecules}. \emph{Chem. Rev.}
  \textbf{2005}, \emph{105}, 4009\relax
\mciteBstWouldAddEndPuncttrue
\mciteSetBstMidEndSepPunct{\mcitedefaultmidpunct}
{\mcitedefaultendpunct}{\mcitedefaultseppunct}\relax
\EndOfBibitem
\bibitem[Maitra \latin{et~al.}(2004)Maitra, Zhang, Cave, and Burke]{burke_tddft_double_ex_jcp_2004}
Maitra,~N.~T.; Zhang,~F.; Cave,~R.~J.; Burke,~K. {Double Excitations 
Within Time-Dependent Density Functional Theory Linear Response}. \emph{J. Chem. Phys.}
  \textbf{2004}, \emph{120}, 5932\relax
\mciteBstWouldAddEndPuncttrue
\mciteSetBstMidEndSepPunct{\mcitedefaultmidpunct}
{\mcitedefaultendpunct}{\mcitedefaultseppunct}\relax
\EndOfBibitem
\bibitem[Besley and Asmuruf(2010)Besley and Asmuruf]{besley_asmuruf_tddft_core_ex_pccp_2010}
Besley,~N.~A.; Asmuruf,~F.~A. {Time-Dependent Density Functional
Theory Calculations of the Spectroscopy of Core Electrons}. \emph{Phys. Chem. Chem. Phys.}
  \textbf{2010}, \emph{12}, 12024\relax
\mciteBstWouldAddEndPuncttrue
\mciteSetBstMidEndSepPunct{\mcitedefaultmidpunct}
{\mcitedefaultendpunct}{\mcitedefaultseppunct}\relax
\EndOfBibitem
\bibitem[Dreuw \latin{et~al.}(2003)Dreuw, Weisman, and
  Head-Gordon]{dreuw_head_gordon_long_range_ct_jcp_2003}
Dreuw,~A.; Weisman,~J.~L.; Head-Gordon,~M. {Long-Range Charge-Transfer Excited
  States in Time-Dependent Density Functional Theory Require Non-Local
  Exchange}. \emph{J. Chem. Phys.} \textbf{2003}, \emph{119}, 2943\relax
\mciteBstWouldAddEndPuncttrue
\mciteSetBstMidEndSepPunct{\mcitedefaultmidpunct}
{\mcitedefaultendpunct}{\mcitedefaultseppunct}\relax
\EndOfBibitem
\bibitem[Dreuw and Head-Gordon(2004)Dreuw, and
  Head-Gordon]{dreuw_head_gordon_long_range_ct_jacs_2004}
Dreuw,~A.; Head-Gordon,~M. {Failure of Time-Dependent Density Functional Theory
  for Long-Range Charge-Transfer Excited States: The
  Zincbacteriochlorin-Bacteriochlorin and Bacteriochlorophyll-Spheroidene
  Complexes}. \emph{J. Am. Chem. Soc.} \textbf{2004}, \emph{126}, 4007\relax
\mciteBstWouldAddEndPuncttrue
\mciteSetBstMidEndSepPunct{\mcitedefaultmidpunct}
{\mcitedefaultendpunct}{\mcitedefaultseppunct}\relax
\EndOfBibitem
\bibitem[Gilbert \latin{et~al.}(2008)Gilbert, Besley, and
  Gill]{delta_scf_mom_gill_jpca_2008}
Gilbert,~A. T.~B.; Besley,~N.~A.; Gill,~P. M.~W. {Self-Consistent Field
  Calculations of Excited States Using the Maximum Overlap Method (MOM)}.
  \emph{{J}. {P}hys. {C}hem. A} \textbf{2008}, \emph{112}, 13164\relax
\mciteBstWouldAddEndPuncttrue
\mciteSetBstMidEndSepPunct{\mcitedefaultmidpunct}
{\mcitedefaultendpunct}{\mcitedefaultseppunct}\relax
\EndOfBibitem
\bibitem[Saunders and Hillier(1973)Saunders, and
  Hillier]{saunders_hillier_level_shift_scf_ijqc_1973}
Saunders,~V.~R.; Hillier,~I.~H. {A ``Level-Shifting" Method for Converging
  Closed Shell Hartree-Fock Wave Functions}. \emph{Int. J. Quantum Chem.}
  \textbf{1973}, \emph{7}, 699\relax
\mciteBstWouldAddEndPuncttrue
\mciteSetBstMidEndSepPunct{\mcitedefaultmidpunct}
{\mcitedefaultendpunct}{\mcitedefaultseppunct}\relax
\EndOfBibitem
\bibitem[Carter-Fenk and Herbert(2020)Carter-Fenk, and
  Herbert]{carter_fenk_herbert_step_scf_jctc_2020}
Carter-Fenk,~K.; Herbert,~J.~M. {State-Targeted Energy Projection: A Simple and
  Robust Approach to Orbital Relaxation of Non-Aufbau Self-Consistent Field
  Solutions}. \emph{J. Chem. Theory Comput.} \textbf{2020}, \emph{16},
  5067\relax
\mciteBstWouldAddEndPuncttrue
\mciteSetBstMidEndSepPunct{\mcitedefaultmidpunct}
{\mcitedefaultendpunct}{\mcitedefaultseppunct}\relax
\EndOfBibitem
\bibitem[Hait and Head-Gordon(2021)Hait, and
  Head-Gordon]{hait_head_gordon_oo_dft_jpcl_2021}
Hait,~D.; Head-Gordon,~M. {Orbital Optimized Density Functional Theory for
  Electronic Excited States}. \emph{J. Phys. Chem. Lett.} \textbf{2021},
  \emph{12}, 4517\relax
\mciteBstWouldAddEndPuncttrue
\mciteSetBstMidEndSepPunct{\mcitedefaultmidpunct}
{\mcitedefaultendpunct}{\mcitedefaultseppunct}\relax
\EndOfBibitem
\bibitem[Gaudoin and Burke(2004)Gaudoin, and
  Burke]{burke_ex_hk_theorem_prl_2004}
Gaudoin,~R.; Burke,~K. {Lack of Hohenberg-Kohn Theorem for Excited States}.
  \emph{Phys. Rev. Lett.} \textbf{2004}, \emph{93}, 173001\relax
\mciteBstWouldAddEndPuncttrue
\mciteSetBstMidEndSepPunct{\mcitedefaultmidpunct}
{\mcitedefaultendpunct}{\mcitedefaultseppunct}\relax
\EndOfBibitem
\bibitem[Eriksen(2020)]{eriksen_decodense_jcp_2020}
Eriksen,~J.~J. {Mean-Field Density Matrix Decompositions}. \emph{{J}. {C}hem.
  {P}hys.} \textbf{2020}, \emph{153}, 214109\relax
\mciteBstWouldAddEndPuncttrue
\mciteSetBstMidEndSepPunct{\mcitedefaultmidpunct}
{\mcitedefaultendpunct}{\mcitedefaultseppunct}\relax
\EndOfBibitem
\bibitem[Peach \latin{et~al.}(2008)Peach, Benfield, Helgaker, and
  Tozer]{helgaker_tozer_ct_lambda_tddft_jcp_2008}
Peach,~M. J.~G.; Benfield,~P.; Helgaker,~T.; Tozer,~D.~J. {Excitation Energies
  in Density Functional Theory: An Evaluation and a Diagnostic Test}. \emph{J.
  Chem. Phys.} \textbf{2008}, \emph{128}, 044118\relax
\mciteBstWouldAddEndPuncttrue
\mciteSetBstMidEndSepPunct{\mcitedefaultmidpunct}
{\mcitedefaultendpunct}{\mcitedefaultseppunct}\relax
\EndOfBibitem
\bibitem[Peach and Tozer(2012)Peach, and Tozer]{peach_tozer_ct_tddft_jpca_2012}
Peach,~M. J.~G.; Tozer,~D.~J. {Overcoming Low Orbital Overlap and Triplet
  Instability Problems in TDDFT}. \emph{J. Phys. Chem. A} \textbf{2012},
  \emph{116}, 9783\relax
\mciteBstWouldAddEndPuncttrue
\mciteSetBstMidEndSepPunct{\mcitedefaultmidpunct}
{\mcitedefaultendpunct}{\mcitedefaultseppunct}\relax
\EndOfBibitem
\bibitem[Guido \latin{et~al.}(2013)Guido, Cortona, Mennucci, and
  Adamo]{mannucci_adamo_ct_tddft_jctc_2013}
Guido,~C.~A.; Cortona,~P.; Mennucci,~B.; Adamo,~C. {On the Metric of Charge
  Transfer Molecular Excitations: A Simple Chemical Descriptor}. \emph{J. Chem.
  Theory Comput.} \textbf{2013}, \emph{9}, 3118\relax
\mciteBstWouldAddEndPuncttrue
\mciteSetBstMidEndSepPunct{\mcitedefaultmidpunct}
{\mcitedefaultendpunct}{\mcitedefaultseppunct}\relax
\EndOfBibitem
\bibitem[Plasser \latin{et~al.}(2014)Plasser, Wormit, and
  Dreuw]{dreuw_ct_tddft_jcp_2014}
Plasser,~F.; Wormit,~M.; Dreuw,~A. {New Tools for the Systematic Analysis and
  Visualization of Electronic Excitations. I. Formalism}. \emph{J. Chem. Phys.}
  \textbf{2014}, \emph{141}, 024106\relax
\mciteBstWouldAddEndPuncttrue
\mciteSetBstMidEndSepPunct{\mcitedefaultmidpunct}
{\mcitedefaultendpunct}{\mcitedefaultseppunct}\relax
\EndOfBibitem
\bibitem[Etienne \latin{et~al.}(2014)Etienne, Assfeld, and
  Monari]{etienne_assfeld_monari_ct_tddft_jctc_2014}
Etienne,~T.; Assfeld,~X.; Monari,~A. {Toward a Quantitative Assessment of
  Electronic Transitions' Charge-Transfer Character}. \emph{J. Chem. Theory
  Comput.} \textbf{2014}, \emph{10}, 3896\relax
\mciteBstWouldAddEndPuncttrue
\mciteSetBstMidEndSepPunct{\mcitedefaultmidpunct}
{\mcitedefaultendpunct}{\mcitedefaultseppunct}\relax
\EndOfBibitem
\bibitem[Moore \latin{et~al.}(2015)Moore, Sun, Govind, Kowalski, and
  Autschbach]{kowalski_autschbach_ct_tddft_jctc_2015}
Moore,~B.,~II; Sun,~H.; Govind,~N.; Kowalski,~K.; Autschbach,~J.
  {Charge-Transfer Versus Charge-Transfer-Like Excitations Revisited}. \emph{J.
  Chem. Theory Comput.} \textbf{2015}, \emph{11}, 3305\relax
\mciteBstWouldAddEndPuncttrue
\mciteSetBstMidEndSepPunct{\mcitedefaultmidpunct}
{\mcitedefaultendpunct}{\mcitedefaultseppunct}\relax
\EndOfBibitem
\bibitem[Krylov(2020)]{krylov_orbital_perspective_jcp_2020}
Krylov,~A.~I. {From Orbitals to Observables and Back}. \emph{J. Chem. Phys.}
  \textbf{2020}, \emph{153}, 080901\relax
\mciteBstWouldAddEndPuncttrue
\mciteSetBstMidEndSepPunct{\mcitedefaultmidpunct}
{\mcitedefaultendpunct}{\mcitedefaultseppunct}\relax
\EndOfBibitem
\bibitem[Eriksen(2021)]{eriksen_local_condensed_phase_jpcl_2021}
Eriksen,~J.~J. {Decomposed Mean-Field Simulations of Local Properties in
  Condensed Phases}. \emph{J. Phys. Chem. Lett.} \textbf{2021}, \emph{16},
  6048\relax
\mciteBstWouldAddEndPuncttrue
\mciteSetBstMidEndSepPunct{\mcitedefaultmidpunct}
{\mcitedefaultendpunct}{\mcitedefaultseppunct}\relax
\EndOfBibitem
\bibitem[Mulliken(1955)]{mulliken_population_jcp_1955}
Mulliken,~R.~S. {Electronic Population Analysis on LCAO-MO Molecular Wave
  Functions. I}. \emph{J. Chem. Phys.} \textbf{1955}, \emph{23}, 1833\relax
\mciteBstWouldAddEndPuncttrue
\mciteSetBstMidEndSepPunct{\mcitedefaultmidpunct}
{\mcitedefaultendpunct}{\mcitedefaultseppunct}\relax
\EndOfBibitem
\bibitem[Knizia(2013)]{knizia_iao_ibo_jctc_2013}
Knizia,~G. {Intrinsic Atomic Orbitals: An Unbiased Bridge Between Quantum
  Theory and Chemical Concepts}. \emph{{J}. {C}hem. {T}heory {C}omput.}
  \textbf{2013}, \emph{9}, 4834\relax
\mciteBstWouldAddEndPuncttrue
\mciteSetBstMidEndSepPunct{\mcitedefaultmidpunct}
{\mcitedefaultendpunct}{\mcitedefaultseppunct}\relax
\EndOfBibitem
\bibitem[Senjean \latin{et~al.}(2021)Senjean, Sen, Repisky, Knizia, and
  Visscher]{knizia_visscher_iao_jctc_2021}
Senjean,~B.; Sen,~S.; Repisky,~M.; Knizia,~G.; Visscher,~L. {Generalization of
  Intrinsic Orbitals to Kramers-Paired Quaternion Spinors, Molecular Fragments
  and Valence Virtual Spinors}. \emph{{J}. {C}hem. {T}heory {C}omput.}
  \textbf{2021}, \emph{17}, 1337\relax
\mciteBstWouldAddEndPuncttrue
\mciteSetBstMidEndSepPunct{\mcitedefaultmidpunct}
{\mcitedefaultendpunct}{\mcitedefaultseppunct}\relax
\EndOfBibitem
\bibitem[Canc{\`e}s \latin{et~al.}(2013)Canc{\`e}s, Maday, and
  Stamm]{cances_maday_stamm_dd_solvation_jcp_2013}
Canc{\`e}s,~E.; Maday,~Y.; Stamm,~B. {Domain Decomposition for Implicit
  Solvation Models}. \emph{J. Chem. Phys.} \textbf{2013}, \emph{139},
  054111\relax
\mciteBstWouldAddEndPuncttrue
\mciteSetBstMidEndSepPunct{\mcitedefaultmidpunct}
{\mcitedefaultendpunct}{\mcitedefaultseppunct}\relax
\EndOfBibitem
\bibitem[Lipparini \latin{et~al.}(2013)Lipparini, Stamm, Canc{\`e}s, Maday, and
  Mennucci]{lipparini_mennucci_dd_cosmo_jctc_2014}
Lipparini,~F.; Stamm,~B.; Canc{\`e}s,~E.; Maday,~Y.; Mennucci,~B. {A Fast
  Domain Decomposition Algorithm for Continuum Solvation Models: Energy and
  First Derivatives}. \emph{J. Chem. Theory Comput.} \textbf{2013}, \emph{9},
  3637\relax
\mciteBstWouldAddEndPuncttrue
\mciteSetBstMidEndSepPunct{\mcitedefaultmidpunct}
{\mcitedefaultendpunct}{\mcitedefaultseppunct}\relax
\EndOfBibitem
\bibitem[Klamt(2017)]{klamt_cosmo_review_wires_2017}
Klamt,~A. {The COSMO and COSMO-RS Solvation Models}. \emph{{W}IREs {C}omput.
  {M}ol. {S}ci.} \textbf{2017}, \emph{8}, e1338\relax
\mciteBstWouldAddEndPuncttrue
\mciteSetBstMidEndSepPunct{\mcitedefaultmidpunct}
{\mcitedefaultendpunct}{\mcitedefaultseppunct}\relax
\EndOfBibitem
\bibitem[Perdew and Schmidt(2001)Perdew, and
  Schmidt]{perdew_jacobs_ladder_aip_conf_proc_2001}
Perdew,~J.~P.; Schmidt,~K. {Jacob's Ladder of Density Functional Approximations
  for the Exchange-Correlation Energy}. \emph{AIP Conf. Proc.} \textbf{2001},
  \emph{577}, 1\relax
\mciteBstWouldAddEndPuncttrue
\mciteSetBstMidEndSepPunct{\mcitedefaultmidpunct}
{\mcitedefaultendpunct}{\mcitedefaultseppunct}\relax
\EndOfBibitem
\bibitem[Nakai(2002)]{nakai_eda_partitioning_cpl_2002}
Nakai,~H. {Energy Density Analysis with Kohn-Sham Orbitals}. \emph{Chem. Phys.
  Lett.} \textbf{2002}, \emph{363}, 73\relax
\mciteBstWouldAddEndPuncttrue
\mciteSetBstMidEndSepPunct{\mcitedefaultmidpunct}
{\mcitedefaultendpunct}{\mcitedefaultseppunct}\relax
\EndOfBibitem
\bibitem[Kikuchi \latin{et~al.}(2009)Kikuchi, Imamura, and
  Nakai]{nakai_eda_partitioning_ijqc_2009}
Kikuchi,~Y.; Imamura,~Y.; Nakai,~H. {One-Body Energy Decomposition Schemes
  Revisited: Assessment of Mulliken-, Grid-, and Conventional Energy Density
  Analyses}. \emph{Int. J. Quantum Chem.} \textbf{2009}, \emph{109}, 2464\relax
\mciteBstWouldAddEndPuncttrue
\mciteSetBstMidEndSepPunct{\mcitedefaultmidpunct}
{\mcitedefaultendpunct}{\mcitedefaultseppunct}\relax
\EndOfBibitem
\bibitem[Not()]{Note-1}
Please note that the sum of the local and global electron-nuclear attractions
  will be identical, whereas this will, in general, not hold true in the case of
  the individual atomic contributions.\relax
\mciteBstWouldAddEndPunctfalse
\mciteSetBstMidEndSepPunct{\mcitedefaultmidpunct}
{}{\mcitedefaultseppunct}\relax
\EndOfBibitem
\bibitem[Becke(1993)]{becke_b3lyp_functional_jcp_1993}
Becke,~A.~D. {Density-Functional Thermochemistry. III. The Role of Exact
  Exchange}. \emph{J. Chem. Phys.} \textbf{1993}, \emph{98}, 5648\relax
\mciteBstWouldAddEndPuncttrue
\mciteSetBstMidEndSepPunct{\mcitedefaultmidpunct}
{\mcitedefaultendpunct}{\mcitedefaultseppunct}\relax
\EndOfBibitem
\bibitem[Stephens \latin{et~al.}(1994)Stephens, Devlin, Chabalowski, and
  Frisch]{frisch_b3lyp_functional_jpc_1994}
Stephens,~P.~J.; Devlin,~F.~J.; Chabalowski,~C.~F.; Frisch,~M.~J. {{\it{Ab
  Initio}} Calculation of Vibrational Absorption and Circular Dichroism Spectra
  Using Density Functional Force Fields}. \emph{J. Phys. Chem.} \textbf{1994},
  \emph{98}, 11623\relax
\mciteBstWouldAddEndPuncttrue
\mciteSetBstMidEndSepPunct{\mcitedefaultmidpunct}
{\mcitedefaultendpunct}{\mcitedefaultseppunct}\relax
\EndOfBibitem
\bibitem[Yanai \latin{et~al.}(2004)Yanai, Tew, and
  Handy]{yanai_tew_handy_camb3lyp_functional_cpl_2004}
Yanai,~T.; Tew,~D.~P.; Handy,~N.~C. {A New Hybrid Exchange-Correlation
  Functional Using the Coulomb-Attenuating Method (CAM-B3LYP)}. \emph{Chem.
  Phys. Lett.} \textbf{2004}, \emph{393}, 51\relax
\mciteBstWouldAddEndPuncttrue
\mciteSetBstMidEndSepPunct{\mcitedefaultmidpunct}
{\mcitedefaultendpunct}{\mcitedefaultseppunct}\relax
\EndOfBibitem
\bibitem[Zhao and Truhlar(2008)Zhao, and
  Truhlar]{zhao_truhlar_m06_functional_tca_2008}
Zhao,~Y.; Truhlar,~D.~G. {The M06 Suite of Density Functionals for Main Group
  Thermochemistry, Thermochemical Kinetics, Noncovalent Interactions, Excited
  States, and Transition Elements: Two New Functionals and Systematic Testing
  of Four M06-Class Functionals and 12 Other Functionals}. \emph{Theor. Chem.
  Acc.} \textbf{2008}, \emph{120}, 215\relax
\mciteBstWouldAddEndPuncttrue
\mciteSetBstMidEndSepPunct{\mcitedefaultmidpunct}
{\mcitedefaultendpunct}{\mcitedefaultseppunct}\relax
\EndOfBibitem
\bibitem[Jensen(2001)]{jensen_pc_basis_sets_jcp_2001}
Jensen,~F. {Polarization Consistent Basis Sets: Principles}. \emph{{J}. {C}hem.
  {P}hys.} \textbf{2001}, \emph{115}, 9113\relax
\mciteBstWouldAddEndPuncttrue
\mciteSetBstMidEndSepPunct{\mcitedefaultmidpunct}
{\mcitedefaultendpunct}{\mcitedefaultseppunct}\relax
\EndOfBibitem
\bibitem[rdk()]{rdkit_prog}
{\texttt{RDKit}}: Open-Source Cheminformatics.
  {\url{https://www.rdkit.org}}\relax
\mciteBstWouldAddEndPuncttrue
\mciteSetBstMidEndSepPunct{\mcitedefaultmidpunct}
{\mcitedefaultendpunct}{\mcitedefaultseppunct}\relax
\EndOfBibitem
\bibitem[Jamorski \latin{et~al.}(2002)Jamorski, Foresman, Thilgen, and
  L{\"u}thi]{luethi_donor_accept_tddft_jcp_2002}
Jamorski,~C.; Foresman,~J.~B.; Thilgen,~C.; L{\"u}thi,~H.-P. {Assessment of
  Time-Dependent Density-Functional Theory for the Calculation of Critical
  Features in the Absorption Spectra of a Series of Aromatic Donor-Acceptor
  Systems}. \emph{J. Chem. Phys.} \textbf{2002}, \emph{116}, 8761\relax
\mciteBstWouldAddEndPuncttrue
\mciteSetBstMidEndSepPunct{\mcitedefaultmidpunct}
{\mcitedefaultendpunct}{\mcitedefaultseppunct}\relax
\EndOfBibitem
\bibitem[Rappoport and Furche(2004)Rappoport, and
  Furche]{furche_dmabn_tddft_jacs_2004}
Rappoport,~D.; Furche,~F. {Photoinduced Intramolecular Charge Transfer in
  4-(dimethyl)aminobenzonitrile --- A Theoretical Perspective}. \emph{J. Am.
  Chem. Soc.} \textbf{2004}, \emph{126}, 1277\relax
\mciteBstWouldAddEndPuncttrue
\mciteSetBstMidEndSepPunct{\mcitedefaultmidpunct}
{\mcitedefaultendpunct}{\mcitedefaultseppunct}\relax
\EndOfBibitem
\bibitem[Loos \latin{et~al.}(2021)Loos, Comin, Blase, and
  Jacquemin]{loos_jacquemin_cipsi_ct_state_jctc_2021}
Loos,~P.-F.; Comin,~M.; Blase,~X.; Jacquemin,~D. {Reference Energies for
  Intramolecular Charge-Transfer Excitations}. \emph{{J}. {C}hem. {T}heory
  {C}omput.} \textbf{2021}, \emph{17}, 3666\relax
\mciteBstWouldAddEndPuncttrue
\mciteSetBstMidEndSepPunct{\mcitedefaultmidpunct}
{\mcitedefaultendpunct}{\mcitedefaultseppunct}\relax
\EndOfBibitem
\bibitem[Yamaguchi \latin{et~al.}(1988)Yamaguchi, Jensen, Dorigo, and
  Houk]{houk_approx_spin_proj_cpl_1988}
Yamaguchi,~K.; Jensen,~F.; Dorigo,~A.; Houk,~K.~N. {A Spin Correction Procedure
  for Unrestricted Hartree-Fock and M{\o}ller-Plesset Wavefunctions for Singlet
  Diradicals and Polyradicals}. \emph{Chem. Phys. Lett.} \textbf{1988},
  \emph{149}, 537\relax
\mciteBstWouldAddEndPuncttrue
\mciteSetBstMidEndSepPunct{\mcitedefaultmidpunct}
{\mcitedefaultendpunct}{\mcitedefaultseppunct}\relax
\EndOfBibitem
\bibitem[Not()]{Note-2}
As in our previous studies (Refs. \citenum{eriksen_decodense_jcp_2020} and
  \citenum{eriksen_local_condensed_phase_jpcl_2021}), a PM localization power
  ($p=2$) has been used to generate the
  IBOs~\cite{lehtola_jonsson_pm_jctc_2014}.\relax
\mciteBstWouldAddEndPunctfalse
\mciteSetBstMidEndSepPunct{\mcitedefaultmidpunct}
{}{\mcitedefaultseppunct}\relax
\EndOfBibitem
\bibitem[Lehtola and J{\'o}nsson(2014)Lehtola, and
  J{\'o}nsson]{lehtola_jonsson_pm_jctc_2014}
Lehtola,~S.; J{\'o}nsson,~H. {Pipek-Mezey Orbital Localization Using Various
  Partial Charge Estimates}. \emph{{J}. {C}hem. {T}heory {C}omput.}
  \textbf{2014}, \emph{10}, 642\relax
\mciteBstWouldAddEndPuncttrue
\mciteSetBstMidEndSepPunct{\mcitedefaultmidpunct}
{\mcitedefaultendpunct}{\mcitedefaultseppunct}\relax
\EndOfBibitem
\bibitem[Foster and Boys(1960)Foster, and Boys]{foster_boys_rev_mod_phys_1960}
Foster,~J.~M.; Boys,~S.~F. {Canonical Configurational Interaction Procedure}.
  \emph{{R}ev. {M}od. {P}hys.} \textbf{1960}, \emph{32}, 300\relax
\mciteBstWouldAddEndPuncttrue
\mciteSetBstMidEndSepPunct{\mcitedefaultmidpunct}
{\mcitedefaultendpunct}{\mcitedefaultseppunct}\relax
\EndOfBibitem
\bibitem[Pipek and Mezey(1989)Pipek, and Mezey]{pipek_mezey_jcp_1989}
Pipek,~J.; Mezey,~P.~G. {A Fast Intrinsic Localization Procedure Applicable for
  {\it{Ab Initio}} and Semiempirical Linear Combination of Atomic Orbital Wave
  Functions}. \emph{{J}. {C}hem. {P}hys.} \textbf{1989}, \emph{90}, 4916\relax
\mciteBstWouldAddEndPuncttrue
\mciteSetBstMidEndSepPunct{\mcitedefaultmidpunct}
{\mcitedefaultendpunct}{\mcitedefaultseppunct}\relax
\EndOfBibitem
\bibitem[Eriksen \latin{et~al.}(2013)Eriksen, Sauer, Mikkelsen, Christiansen,
  Jensen, and Kongsted]{eriksen_pna_tddft_mol_phys_2013}
Eriksen,~J.~J.; Sauer,~S. P.~A.; Mikkelsen,~K.~V.; Christiansen,~O.;
  Jensen,~H.-J.~A.; Kongsted,~J. {Failures of TDDFT in Describing the Lowest
  Intramolecular Charge-Transfer Excitation in {\textit{para}}-Nitroaniline}.
  \emph{{M}ol. {P}hys.} \textbf{2013}, \emph{111}, 1235\relax
\mciteBstWouldAddEndPuncttrue
\mciteSetBstMidEndSepPunct{\mcitedefaultmidpunct}
{\mcitedefaultendpunct}{\mcitedefaultseppunct}\relax
\EndOfBibitem
\bibitem[Grabowski \latin{et~al.}(2003)Grabowski, Rotkiewicz, and
  Rettig]{grabowski_rettig_dmabn_chem_rev_2003}
Grabowski,~Z.~R.; Rotkiewicz,~K.; Rettig,~W. {Structural Changes Accompanying
  Intramolecular Electron Transfer: Focus on Twisted Intramolecular
  Charge-Transfer States and Structures}. \emph{Chem. Rev.} \textbf{2003},
  \emph{103}, 3899\relax
\mciteBstWouldAddEndPuncttrue
\mciteSetBstMidEndSepPunct{\mcitedefaultmidpunct}
{\mcitedefaultendpunct}{\mcitedefaultseppunct}\relax
\EndOfBibitem
\bibitem[Uoyama \latin{et~al.}(2012)Uoyama, Goushi, Shizu, Nomura, and
  Adachi]{adachi_tadf_nature_2012}
Uoyama,~H.; Goushi,~K.; Shizu,~K.; Nomura,~H.; Adachi,~C. {Highly Efficient
  Organic Light-Emitting Diodes from Delayed Fluorescence}. \emph{Nature}
  \textbf{2012}, \emph{492}, 234\relax
\mciteBstWouldAddEndPuncttrue
\mciteSetBstMidEndSepPunct{\mcitedefaultmidpunct}
{\mcitedefaultendpunct}{\mcitedefaultseppunct}\relax
\EndOfBibitem
\bibitem[Hait \latin{et~al.}(2016)Hait, Zhu, McMahon, and
  Van~Voorhis]{hait_voorhis_tadf_jctc_2016}
Hait,~D.; Zhu,~T.; McMahon,~D.~P.; Van~Voorhis,~T. {Prediction of Excited-State
  Energies and Singlet-Triplet Gaps of Charge-Transfer States Using a
  Restricted Open-Shell Kohn-Sham Approach}. \emph{J. Chem. Theory Comput.}
  \textbf{2016}, \emph{12}, 3353\relax
\mciteBstWouldAddEndPuncttrue
\mciteSetBstMidEndSepPunct{\mcitedefaultmidpunct}
{\mcitedefaultendpunct}{\mcitedefaultseppunct}\relax
\EndOfBibitem
\bibitem[de~Silva(2019)]{de_silva_tadf_jpcl_2019}
de~Silva,~P. {Inverted Singlet-Triplet Gaps and Their Relevance to Thermally
  Activated Delayed Fluorescence}. \emph{J. Phys. Chem. Lett.} \textbf{2019},
  \emph{10}, 5674\relax
\mciteBstWouldAddEndPuncttrue
\mciteSetBstMidEndSepPunct{\mcitedefaultmidpunct}
{\mcitedefaultendpunct}{\mcitedefaultseppunct}\relax
\EndOfBibitem
\bibitem[Casanova(2018)]{casanova_sf_chem_rev_2018}
Casanova,~D. {Theoretical Modeling of Singlet Fission}. \emph{Chem. Rev.}
  \textbf{2018}, \emph{118}, 7164\relax
\mciteBstWouldAddEndPuncttrue
\mciteSetBstMidEndSepPunct{\mcitedefaultmidpunct}
{\mcitedefaultendpunct}{\mcitedefaultseppunct}\relax
\EndOfBibitem
\bibitem[Fumanal and Corminboeuf(2020)Fumanal, and
  Corminboeuf]{corminboeuf_sf_jpcl_2020}
Fumanal,~M.; Corminboeuf,~C. {Direct, Mediated, and Delayed Intramolecular
  Singlet Fission Mechanism in Donor-Acceptor Copolymers}. \emph{J. Phys. Chem.
  Lett.} \textbf{2020}, \emph{11}, 9788\relax
\mciteBstWouldAddEndPuncttrue
\mciteSetBstMidEndSepPunct{\mcitedefaultmidpunct}
{\mcitedefaultendpunct}{\mcitedefaultseppunct}\relax
\EndOfBibitem
\bibitem[Kaduk \latin{et~al.}(2012)Kaduk, Kowalczyk, and
  Van~Voorhis]{van_voorhis_const_dft_chem_rev_2012}
Kaduk,~B.; Kowalczyk,~T.; Van~Voorhis,~T. {Constrained Density Functional
  Theory}. \emph{Chem. Rev.} \textbf{2012}, \emph{112}, 321\relax
\mciteBstWouldAddEndPuncttrue
\mciteSetBstMidEndSepPunct{\mcitedefaultmidpunct}
{\mcitedefaultendpunct}{\mcitedefaultseppunct}\relax
\EndOfBibitem
\end{mcitethebibliography}
\end{document}